\documentclass[12pt,a4paper]{elsart}
\usepackage{epsfig}
\usepackage{graphics}
\begin{document}
\begin{frontmatter}
\title{Measurement of $e^+e^- \to\phi\to K^+ K^-$
cross section with CMD-2 detector at VEPP-2M Collider}
\author[BINP]{R.R.Akhmetshin},
\author[BINP,NSU]{V.M.Aulchenko},
\author[BINP]{V.Sh.Banzarov},
\author[BINP]{L.M.Barkov},
\author[BINP]{S.E.Baru},
\author[BINP]{N.S.Bashtovoy},
\author[BINP,NSU]{A.E.Bondar},
\author[BINP]{D.V.Bondarev},
\author[BINP]{A.V.Bragin},
\author[BINP,NSU]{S.I.Eidelman},
\author[BINP,NSU]{D.A.Epifanov},
\author[BINP,NSU]{G.V.Fedotovich},
\author[BINP]{N.I.Gabyshev},
\author[BINP,NSU]{A.A.Grebeniuk},
\author[BINP,NSU]{D.N.Grigoriev},
\author[BINP,NSU]{F.V.Ignatov},
\author[BINP]{S.V.Karpov},
\author[BINP,NSU]{V.F.Kazanin},
\author[BINP,NSU]{B.I.Khazin},
\author[BINP,NSU]{I.A.Koop},
\author[BINP,NSU]{P.P.Krokovny},
\author[BINP,NSU]{A.S.Kuzmin},
\author[BINP]{I.B.Logashenko},
\author[BINP]{P.A.Lukin\thanksref{someone}},
\author[BINP]{A.P.Lysenko},
\author[BINP]{K.Yu.Mikhailov},
\author[BINP]{V.S.Okhapkin},
\author[BINP,NSU]{E.A. Perevedentsev},
\author[BINP,NSU]{A.S.Popov},
\author[BINP]{S.I. Redin},
\author[BINP]{A.A.Ruban},
\author[BINP]{N.M.Ryskulov},
\author[BINP]{Yu.M.Shatunov},
\author[BINP,NSU]{B.A.Shwartz},
\author[BINP,NSU]{A.L.Sibidanov},
\author[BINP]{I.G.Snopkov},
\author[BINP,NSU]{E.P.Solodov},
\author[BINP]{Yu.V.Yudin}

\address[BINP]{Budker Institute of Nuclear Physics, Novosibirsk,
630090, Russia}
\address[NSU]{Novosibirsk State University, Novosibirsk, 630090,
Russia}
%\address[YALE]{Yale University, New Haven, CT 06511, USA}
%\address[BU]{Boston University, Boston, MA 02215, USA}
%\address[PITT]{University of Pittsburgh, Pittsburgh, PA, 15260, USA}
\thanks[someone]{contact person. e-mail:P.A.Lukin@inp.nsk.su}

\begin{abstract}
The process  $e^+e^- \to\phi\to K^+ K^-$
has been studied  with the CMD-2 detector  using about 542 000
events detected  in the center-of-mass
energy range from 1.01 to 1.034~GeV. The systematic error of the
cross section is estimated to be 2.2\%. The $\phi(1020)$ meson
parameters in the $\phi\to K^+K^-$ decay channel have been measured:
$\sigma_0(\phi\to K^+K^-) = 2016\pm8\pm 44$ nb,
$m_{\phi} = 1019.441\pm 0.008\pm 0.080$ MeV/c$^2$,
$\Gamma_{\phi} = 4.24\pm 0.02\pm 0.03$ MeV,
$B_{e^+e^-}B_{K^+K^-} = (14.27\pm 0.05\pm 0.31)\times 10^{-5}$.
\end{abstract}
%%%%%
\end{frontmatter}
\section{Introduction}
\hspace*{\parindent} A study of the process $e^+e^-\to K^+K^-$
is of interest for a number of physical problems.
Since the $K^+K^-$ final state is the main $\phi(1020)$ meson
decay channel,
the resonance parameters can be obtained
by measuring  the cross section of the process in the energy
range around the $\phi(1020)$ meson mass~\cite{CMD-295,SND01}.
The isovector part
of the $e^+e^-\to K\bar{K}$ cross section (both
$K^+K^-$ and $K^0_LK^0_S$ final states should be considered)
can be related to the $\tau^- \to K^-K^0\nu_{\tau}$  decay by using
conservation of vector current (CVC)\cite{cvc}.
%Assuming the hypothesis of
%the factorization it can be also used to account for the production of
%the kaon pairs in the $B^- \to D^0 K^- K^0$ decays \cite{dru}.
Finally, the process under study
is used in the calculation of the hadronic contribution to
the muon anomalous magnetic moment~\cite{g-21}. In view of the
increasing experimental accuracy in the measurement of this
quantity \cite{g-22}, any significant contribution like that from the
process $e^+e^-\to K^+K^-$ should be measured with
adequate precision.

At the energy around the $\phi(1020)$ meson
mass  low momenta kaons from the process $e^+e^-\to K^+K^-$
have large probabilities for a nuclear interaction,
decays in flight and kaon stop in a thin layer of the detector material.
That introduces large uncertainties in the detection efficiency and
increases systematic errors in the cross section.
Earlier measurement of the cross section performed by the
CMD-2 collaboration~\cite{CMD-295} at the VEPP-2M collider~\cite{VEPP-2M},
was based on a relatively small data
sample and had a  systematic accuracy about 4\%.
The SND collaboration~\cite{SND01}
%also at the VEPP-2M collider~\cite{VEPP-2M},
used significantly larger statistics to
study the reaction $e^+e^-\to K^+K^-$.
The experiment was based on
the integrated luminosity of 8.5~pb$^{-1}$, but the accuracy
of the cross section was limited by
systematic errors estimated to be 7.1\%.

In this work we report a  measurement of the
$e^+e^-\to K^+K^-$ cross section
based on 1.0 pb$^{-1}$ of data collected with the CMD-2 Detector
\cite{CMD-2} at the VEPP-2M collider from 1.01 to 1.034~GeV center-of-mass
($E_{c.m.}=\sqrt{s}$) energy.
A special procedure to extract the detection efficiency from data is developed
and the systematic uncertainty on the
cross section is estimated to be 2.2\%.
\section{Detector and experiment}
\hspace*{\parindent} The CMD-2 detector
 has been described in detail elsewhere~\cite{CMD-2}.
The detector tracking system  consists of the cylindrical
drift chamber (DC)~\cite{DC} surrounding the  interaction point,
and proportional Z-chamber (ZC)~\cite{ZC} for a precise
measurement of polar angles, both also used as a charged trigger.
Both chambers are inside a
thin (0.38~$X_0$) superconducting solenoid~\cite{HFIELD} with a field of 1~T.
The barrel electromagnetic calorimeter~\cite{CsI} is
placed outside the solenoid and consists of
892 CsI crystals.
The muon-range system~\cite{MU} of
the detector, also located outside the solenoid, is based on streamer tubes.
The end-cap electromagnetic calorimeter~\cite{BGO} based
on the 680 BGO crystals makes the detector almost hermetic for photons.
In this experiment we require a charged-trigger signal from at least one
charged track and any ($>$20 MeV) energy deposition in the barrel
electromagnetic calorimeter.

The data sample used in the analysis was collected in two scans of
the center-of-mass energy range 1.01~--~1.034~GeV. In the scans
the beam energy was increased
from 505~MeV to 517~MeV with a 0.5~MeV step.
To determine the detection efficiency,
we simulated 50000 events~\cite{cmd2sim} of the process
$e^+e^-\to K^+K^-(\gamma)$ at each energy point.
%, while in the second one
%it was increased from 505~MeV to 514~MeV with the same energy
%step.
\section{Event selection}
\begin{figure}[tbh]
\includegraphics[width=0.48\textwidth]{select_11.eps}\hfill
\includegraphics[width=0.48\textwidth]{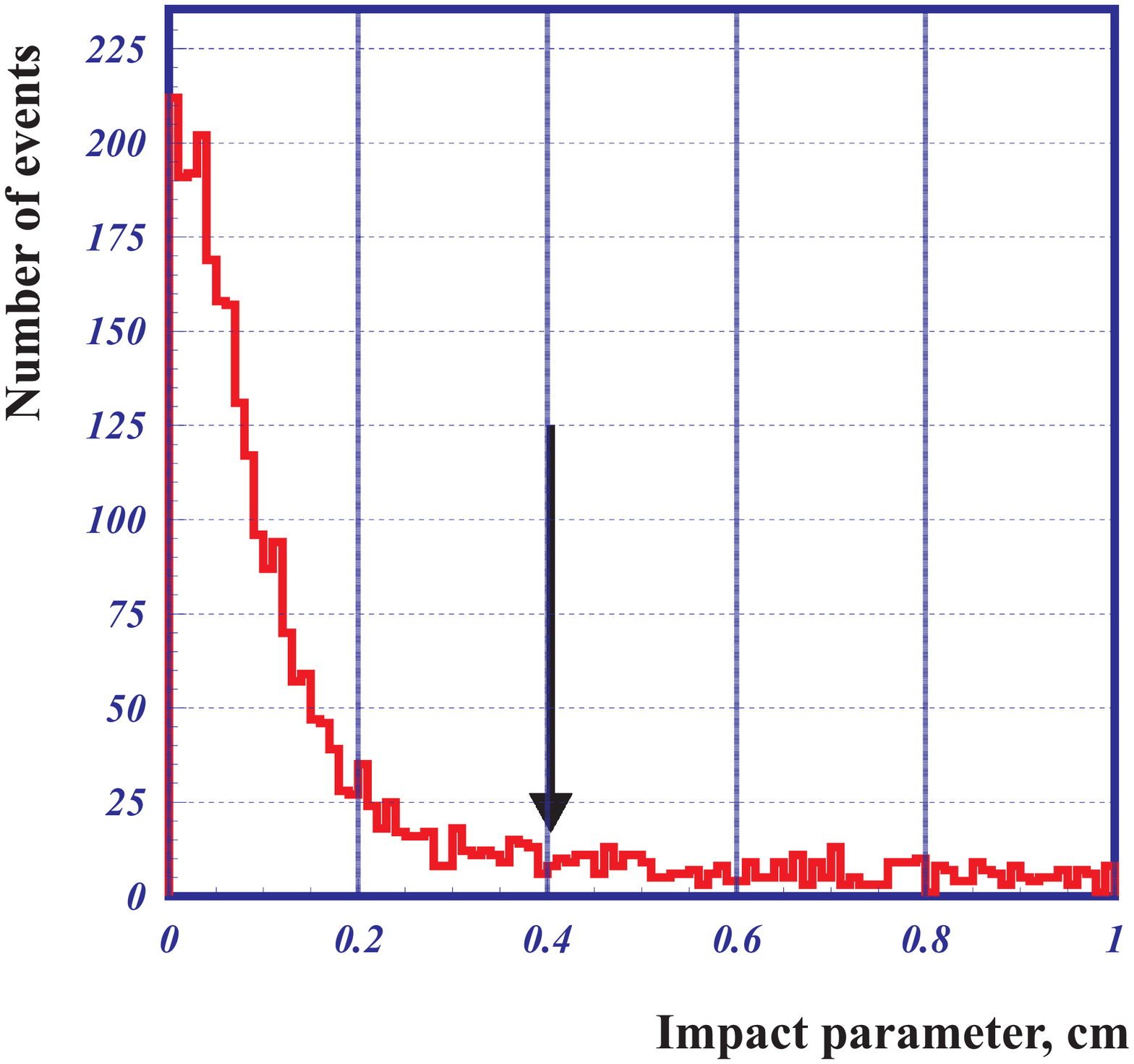}\\
\parbox[t]{0.48\textwidth}{
\caption{\label{sel1}The track ionization losses versus track momentum}}\hfill
\parbox[t]{0.48\textwidth}{
\caption{\label{sel2}The distribution of the track impact parameter in
the $R-\varphi$ plane.}}
\end{figure}

\hspace*{\parindent} A candidate to a $e^+e^-\to K^+K^-$ event is an
event with two low-momentum tracks and high ionization losses,
originating from the interaction region. There is a number of effects
leading to the loss of a charged-kaon track: decays in flight,
nuclear interactions, track reconstruction inefficiency etc. If one
track is not reconstructed, the event can still be identified using
a second detected track. Using single-track events to study detection
efficiency we can significantly reduce various systematic errors.
In our analysis we select events with one or two
``good kaons'' found, where a ``good kaon'' is defined according to
the following criteria:
\begin{itemize}
\item Track polar angle is $1.0 < \theta_{K} < \pi - 1.0$ radians
\item Track total momentum is $P_{\rm tot} < 200 $ MeV/$c$
\item Track ionization loss is $dE/dx > 2\cdot dE/dx_{\rm MIP} $
\item Track impact parameter in $R-\varphi$ plane is $\rho < 0.4$ cm.
\end{itemize}
Figure ~\ref{sel1} shows a scatter plot of the track
ionization losses vs. track total momentum  for all two-track events.
Lines show the boundaries of applied
selections which allow to separate events with charged kaon(s) from
other reactions. The distribution
of the track impact parameter in the $R-\varphi$ plane is shown in
Fig.~\ref{sel2} for the remaining events.
The vertical arrow shows the applied selection.

\begin{figure}[tbh]
\includegraphics[width=0.48\textwidth]{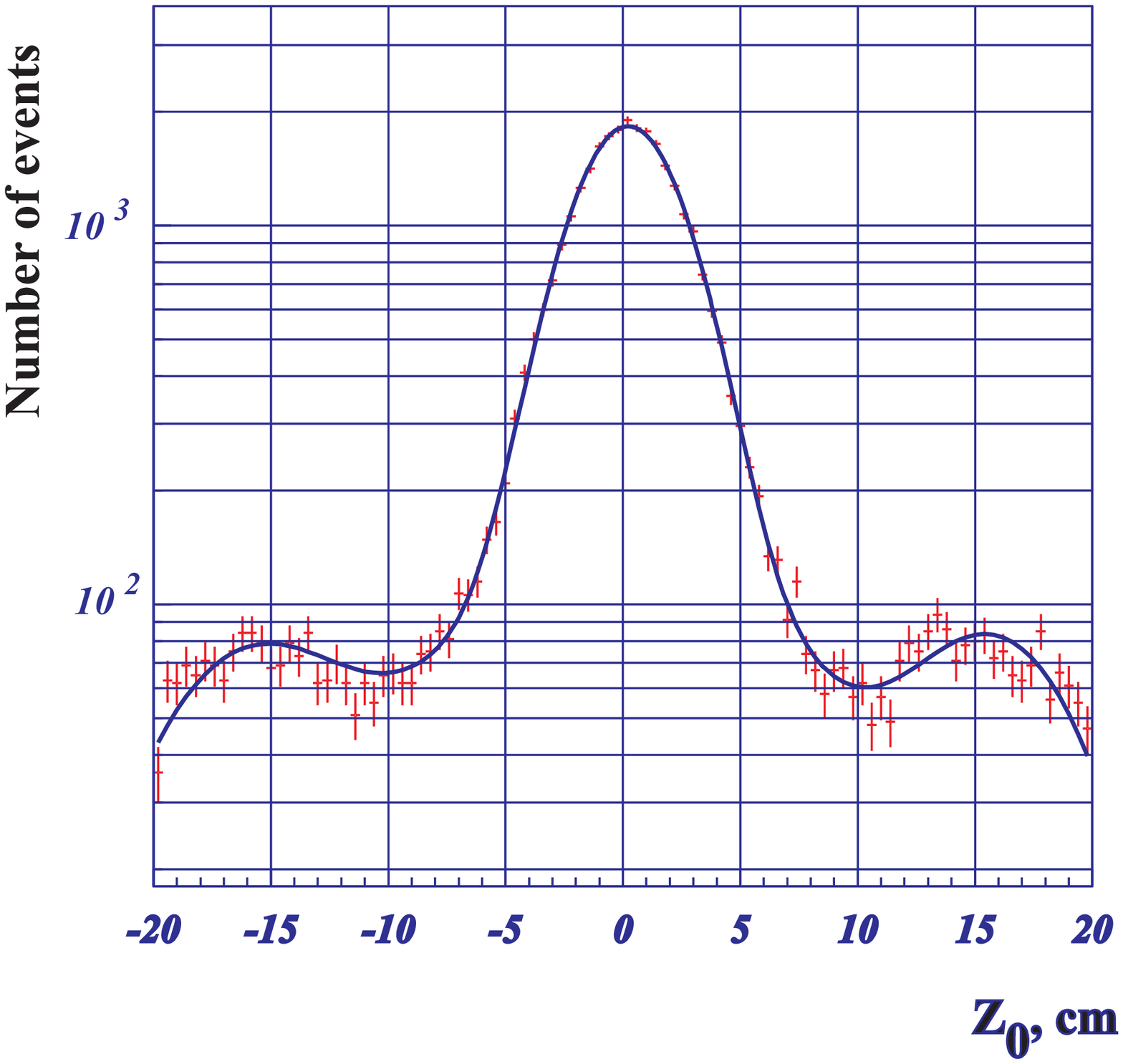}\hfill
\includegraphics[width=0.48\textwidth]{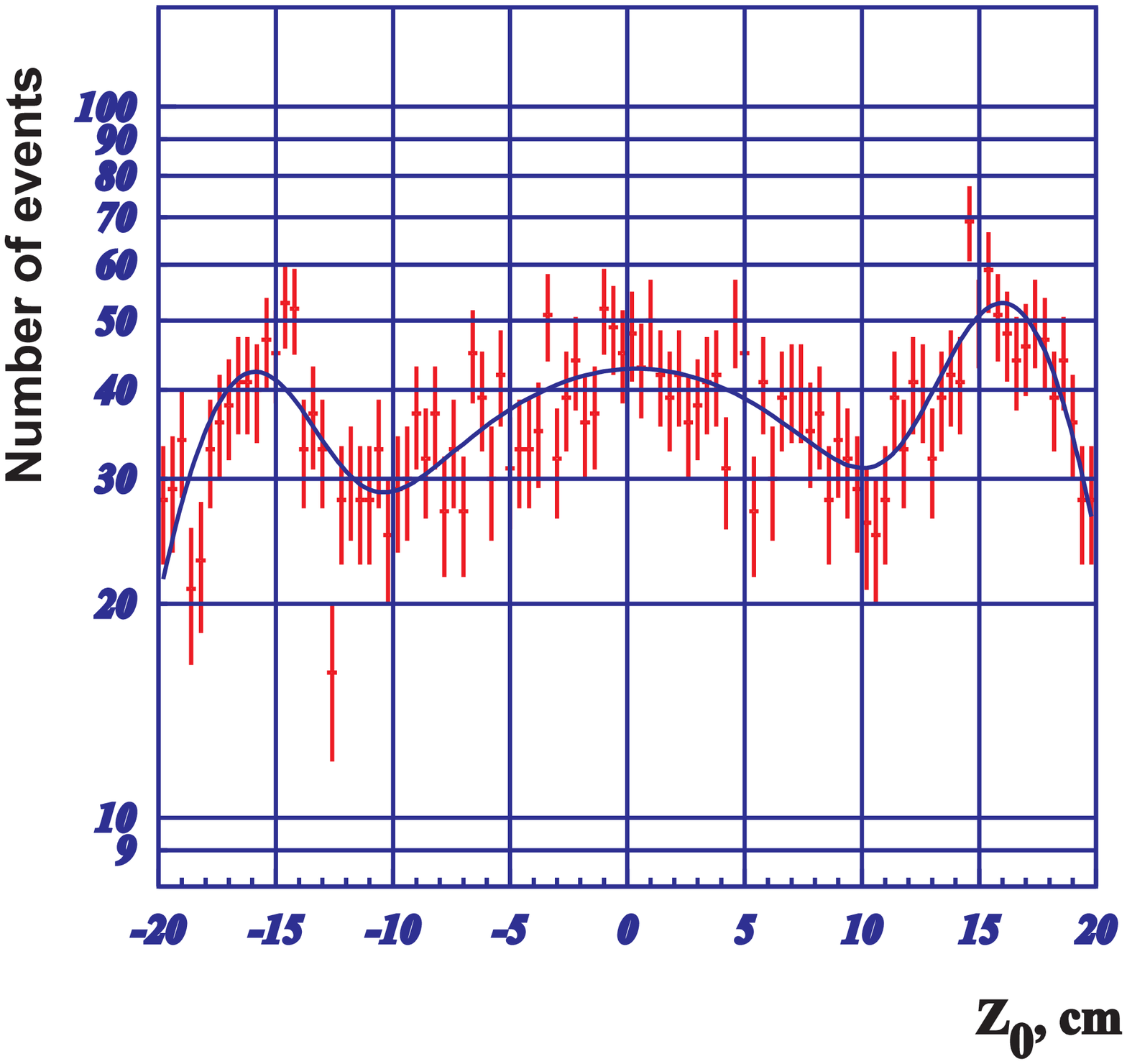}\\
\parbox[t]{0.48\textwidth}{\caption{\label{z0}The distribution of the
Z-coordinate of the point closest to the interaction region for events
with one ``good'' kaon. The curve shows the result of the fit
described in the text.}}\hfill
\parbox[t]{0.48\textwidth}{\caption{\label{z0bkg}
The distribution of the Z-coordinate of the point closest to the interaction
region for background events. The curve shows the results of the fit
described in the text.}}
\end{figure}

The number of events with one or two ``good'' kaons found is
determined from the distribution of a Z-coordinate of the point
closest to the interaction region along the beam axis.
Figure~\ref{z0}
demonstrates the Z-coordinate distribution for
events with one ``good kaon''.
A background from the beam-gas and beam-pipe interactions
producing low-momentum protons or pions is clearly seen.
This background contributes about 15\% to a sample of
one ``good'' kaon events and is significantly
smaller (0.4\%)  if both tracks are identified as
``good kaons''.

To extract the number of signal events, the distribution is fitted
to a  sum of a Gaussian function describing the
interaction region and a smooth function
describing the background. The shape of the background distribution
was derived from the
analysis of events collected at the energy point below the threshold
of charged kaon pair production. The Z-coordinate distribution
obtained at the center-of-mass energy of 0.984~GeV
is shown in Fig.~\ref{z0bkg} and is fitted to a sum of three Gaussian
functions. The obtained
values of the fitting parameters but the number of background events
are then used for the background description at each energy point.
The background is relatively small, so that variations of these parameters
or the alternative description of the background with the ``flat''
distribution do not change the final number of signal events by
more than 0.4\%.

%At $\phi$ meson peak the ratio of number of backround events with one ``good kaon'' found
%to the sum of numbers of signal events with one and two ``good kaon(s)'' is estimated to be
%about 6\%. The ratio of number of background events with two ``good kaons'' found to the
%sum of numbers of signal events with one and two ``good kaons'' found is about 0.4\%.

After background subtraction we select 178932$\pm$432
events with one ``good kaon'' and
363490$\pm$604 events with
two ``good kaons''. The number of selected events for each energy point is
presented in Table~\ref{tab}. By varying the selection criteria we
estimate the systematic error on these numbers as 1.4\%.

%Varying of the parameters of function, described distribution of the background events, within
%$\pm$10\% from its values leads to varying of the cross section within 0.4\%. We estimate our
%systematic error due to background subtraction by this value and quadratically added it the systematic
%error due to selection criteria.
%
\section{Cross section}
\hspace*{\parindent}At each energy point the $e^+e^-\to K^+K^-$ cross
section is calculated according to the formula:
\begin{equation}
\sigma =
\frac{N_1+N_2}{\varepsilon\mathcal{L}\cdot(1+\delta_{\rm{rad}})},
%\cdot\frac{1+\Delta_{\rm{EXP}}}{1+\Delta_{\rm{SIM}}},
\label{xs}
\end{equation}
where $N_1$ and $N_2$ are the numbers of events with one or two
``good'' kaons,  $\varepsilon$ is the detection
efficiency obtained from the MC simulation~\cite{cmd2sim} with some
corrections from data,
$\mathcal{L}$ is the integrated luminosity calculated
with a 1\% accuracy using events of large angle Bhabha
scattering~\cite{lum} and
$(1+\delta_{\rm{rad}})$ is the correction for initial-state radiation
determined  with a 0.5\% accuracy according to Ref.~\cite{rad}.

The detection efficiency is determined from the following formula:
\begin{equation}
\varepsilon = \varepsilon_{\rm{geom}}\cdot\varepsilon_{\rm{TF}}\cdot
\varepsilon_{\rm{CsI}}
\cdot\frac{1+\Delta_{\rm{SIM}}}{1+\Delta_{\rm{EXP}}},
\label{eff}
\end{equation}
where the acceptance $\varepsilon_{\rm{geom}}$ is calculated
as the ratio of the number of events passing the selection criteria to
the initial number of MC simulated events,
$\varepsilon_{\rm{TF}}\cdot\varepsilon_{\rm{CsI}}$ is
the product of the charged-trigger efficiency and a probability to have
energy deposition in the CsI calorimeter.

The number of events with one ``good'' kaon is about 50\% of that with
two ``good'' kaons. Therefore, using the sum of events with one and two ``good''
kaons we increase the detection efficiency and decrease
the uncertainty due to an incorrect description of the track losses
in the MC simulation.
%The typical value of the acceptance is $\varepsilon_{\rm{geom}} = 0.640\pm 0.001$.
In Eq.~\ref{eff} we introduce
$\Delta_{\rm{EXP}}$ and $\Delta_{\rm{SIM}}$ as the corrections
describing the losses of both charged kaons for experimental
and MC simulated events, respectively, taking into account
a different probability of nuclear interaction as well as
the different number of hits for tracks of
positive and negative kaons. We found  no statistically significant
difference in the losses of positive and negative kaons due to the
effects mentioned above.
For example, from the $N_1$ and $N_2$ values in Table~\ref{tab}
at $\sqrt{s} = 1020.1$ MeV and assuming no correlations, we estimate
a probability to lose both kaons to be $\Delta_{\rm{EXP}}=0.035$ in good
agreement with the MC simulation. The difference in $\Delta_{\rm{EXP}}$
and $\Delta_{\rm{SIM}}$ at all
energy points does not exceed 0.7\% and this value is taken as
an estimate of the systematic error in the acceptance calculation.

The charged-trigger efficiency ($\varepsilon_{\rm TF}$)
was estimated to be $0.920\pm0.003$ using a special computer code simulating
CMD-2 trigger~\cite{ZC}. The difference between the trigger efficiency
for experimental and Monte Carlo events is 1\% and is used as an
estimate of the corresponding systematic error.

The positive trigger decision also requires the presence of
at least one cluster in the CsI calorimeter with the energy deposition
greater than 20 MeV. The efficiency $\varepsilon_{\rm{CsI}}$ is calculated
in a similar way and is $0.970\pm0.001$.

The total calculated efficiency for each energy point is listed in
Table~\ref{tab}. The beam energy at each point was
determined using a procedure described in detail in~\cite{lum}.

\begin{table}
\caption{\label{tab}The number of events, integrated luminosity, detection efficiency,
radiative correction, cross section of the $e^+e^-\to\phi\to K^+K^-$ process.}
\vspace*{0.5cm}
\begin{center}
\hspace*{-5mm}\begin{tabular}{ccccccc}
\hline
$\sqrt{s}$, MeV & $N_{2tr}$             & $N_{1tr}$   & $\mathcal{L}$, nb$^{-1}$& 
$\varepsilon$   & $(1+\delta_{rad})$    &$\sigma$, nb \\
\hline
 \multicolumn{7}{c}{First scan}\\
\hline
1011.57$\pm$0.26&986$\pm$31&679$\pm$33&51.3&0.528&0.733&  83.57$\pm$5.89\\   
1016.12$\pm$0.03&8762$\pm$94    &  4735$\pm$  77    &  60.9&0.561&0.712& 549.20$\pm$7.47\\    
1017.02$\pm$0.02&15103$\pm$123    &  7872$\pm$  99    &66.3&0.568&0.707& 853.45$\pm$10.33\\    
1017.97$\pm$0.02&37019$\pm$193    &  19121$\pm$  152    &98.2&0.581&0.705&1389.78$\pm$12.64\\    
1019.20$\pm$0.02&56647$\pm$239    &  28681$\pm$  184    &101.5&0.582&0.725&2020.04$\pm$11.94\\
1020.10$\pm$0.02&53382$\pm$232    &  25265$\pm$  173    &  97.2&0.585&0.763&1825.51$\pm$13.83\\    
1020.97$\pm$0.02&39848$\pm$200    &  17920$\pm$  146    &89.7&0.590&0.816&1333.87$\pm$13.12\\    
1021.81$\pm$0.03&19327$\pm$139    &  8307$\pm$  101    &  57.9&0.593&0.871&917.57$\pm$12.46\\    
1022.75$\pm$0.05&6874$\pm$83    &  3518$\pm$  65    &29.6&0.597&0.934& 626.16$\pm$15.11\\    
1028.33$\pm$0.26&1569$\pm$40    &  815$\pm$  36    &  21.3&0.630&1.241&  143.80$\pm$10.40\\    
1034.06$\pm$0.26&1153$\pm$34    &  550$\pm$  35    & 26.1&0.644&1.452&70.50$\pm$7.54 \\    
\hline
 \multicolumn{7}{c}{Second scan}\\
\hline
   1011.36$\pm$0.26&     539$\pm$23 &  394$\pm$25 &31.9&0.524&0.734&       75.72$\pm$       5.82\\    
   1016.02$\pm$0.03&    1328$\pm$37 &  659$\pm$29 & 9.8&0.563&0.712&       503.76$\pm$       10.77\\    
   1017.09$\pm$0.02&   11917$\pm$110& 6015$\pm$85 &49.6&0.570&0.707&       889.08$\pm$       10.73\\    
   1018.02$\pm$0.05&   24649$\pm$157&11892$\pm$118&62.2&0.584&0.706&       1423.23$\pm$       25.52\\    
   1018.89$\pm$0.02&    8766$\pm$94 & 3946$\pm$68 &15.7&0.585&0.716&       1951.23$\pm$       24.98 \\   
   1019.68$\pm$0.02&   43494$\pm$209&20534$\pm$154&76.1&0.581&0.743&       1971.87$\pm$       12.51\\    
   1020.72$\pm$0.03&   16900$\pm$130& 9671$\pm$105&40.0&0.579&0.800&       1435.03$\pm$       18.17\\    
   1021.74$\pm$0.03&   12290$\pm$111& 7086$\pm$91 &40.6&0.585&0.867&       933.69$\pm$       14.43 \\   
   1022.67$\pm$0.04&    2435$\pm$49 & 1035$\pm$36 &10.2&0.603&0.929&       606.55$\pm$       16.99\\    
   1028.58$\pm$0.03&     502$\pm$22 &  247$\pm$19 & 6.0&0.633&1.252&       158.10$\pm$       12.16\\    
\hline
\end{tabular}
\end{center}
\end{table}

The total systematic error on the cross section is estimated to
be equal to 2.2\% obtained by adding in quadrature contributions
from various sources listed in Table~\ref{syst}.
%We assume that all sources of the systematic
%error are completely correlated and our estimation of the systematic
%error is conservative one.
%
\begin{table}
\caption{\label{syst}Contributions to the systematic error of
the $e^+e^-\to K^+K^-$ cross section}
\hspace*{5mm}
\begin{center}
\begin{tabular}{|c|c|}
\hline
Source & Contribution,\% \\
\hline
\hline
Selection criteria   & 1.4 \\
Trigger efficiency   & 1.0 \\
Luminosity           & 1.0 \\
Acceptance           & 0.7 \\
Radiative correction & 0.5 \\
Background shape     & 0.4 \\
\hline
Total       & 2.2 \\
\hline
\end{tabular}
\end{center}
\end{table}

The experimental points are fit with the Breit-Wigner
function~\cite{SND01}, which includes the contributions from the $\rho$,
$\omega$ and $\phi$ mesons:
%\begin{eqnarray}
%\sigma(s) & = & \frac{1}{s^{5/2}}\cdot\frac{q^3(s)}{q^3(m^2_{\phi})}\cdot
%\left|\frac{\Gamma_{\phi}m^3_{\phi}\sqrt{\sigma(\phi\to K^+K^-)m_{\phi}}}
%{D_{\phi}(s)}e^{\imath\psi_{KK}} -\right.\nonumber\\
% &-& \frac{\sqrt{\Gamma_{\phi}\Gamma_{\omega}m^2_{\phi}m^3_{\omega}6\pi
% B(\omega\to e^+e^-)B(\phi\to K^+K^-)}}{D_{\omega}(s)} - \nonumber\\
% &-&\left. \frac{\sqrt{\Gamma_{\phi}\Gamma_{\rho}m^2_{\phi}m^3_{\rho}6\pi
%  B(\rho\to e^+e^-)B(\phi\to K^+K^-)}}{D_{\rho}(s)}\right|^2\frac{Z(s)}
%{Z(m^2_{\phi})}\nonumber,
%\end{eqnarray}
%where $\sigma(\phi\to K^+K^-)$ is the $\phi$ meson peak cross section,
%$q(s) = \sqrt{s/4-m^2_{K^{\pm}}}$ is the charged kaon momentum,
%$D_V(s)=(m^2_V-s-\imath\sqrt{s}\Gamma_V(s))$ is the propagator of the
%vector meson $V$, $\psi_{KK}$ is the phase of $\phi$-$\rho$ mesons interference
%and we set it to $\pi$ according to
%a naive quark model.
%$$
%Z(s)  =  1+\alpha\cdot\pi\cdot\frac{1+v^2}{2\cdot v},
%$$
%$$
%v     =  \sqrt{1-\frac{m^2_{K^{\pm}}}{s}}
%$$
%describes charged kaons Colomb interaction in final state.~\cite{Colomb}
\begin{eqnarray}
\sigma(s)_{K^+K^-} & = & \frac{8\pi\alpha}{3\cdot s^{5/2}}\cdot q^3(s)\left |\sum_V\frac{g_{V\gamma}g_{VK^+K^-}}{D_V(s)}
e^{\imath\psi_{V}} + A_{K^+K^-}\right|^2\frac{Z(s)}{Z(m^2_{\phi})},
\end{eqnarray}
where V means $\rho(770),\omega(782)$ or $\phi(1020)$ mesons. $q(s) = \sqrt{s/4-m^2_{K^{\pm}}}$
is the charged kaon momentum, $D_V(s)=(m^2_V-s-\imath\sqrt{s}\Gamma_V(s))$ is the propagator of the vector meson $V$,
$g_{V\gamma}$ is a constant describing the  coupling of the
meson $V$ with a photon and $g_{VKK}$~---~coupling constant of the
meson $V$ with a $K^+K^-$ pair,
$\psi_V$ is the phase.
The coupling constants $g_{V\gamma} g_{VKK}$ are related to the
product of the branching fractions $B(V\to e^+e^-)B(V\to K^+K^-)$
according to:
\begin{eqnarray}
g_{V\gamma}g_{VKK} & = & 3m^2_V\Gamma_V\sqrt{\frac{m_V B(V\to e^+e^-)B(V\to K^+K^-)}{\alpha q^3(m^2_V)}}.
\end{eqnarray}
Since the $\rho(770)$ and $\omega(782)$ mesons are below the
$K^+K^-$ pair production threshold, we calculate
$B(\rho,\omega\to K^+K^-)$ using the corresponding relation
from simple quark model (see, for example, Ref.~\cite{NuclPhys}):
\begin{eqnarray}
\frac{|g_{\rho KK}|}{|g_{\phi KK}|} & = & \frac{|g_{\omega KK}|}{|g_{\phi KK}|}  =  \frac{1}{\sqrt{2}}.
\nonumber
\end{eqnarray}
The phases $\psi_{\phi}$ and $\psi_{\omega}$ are equal to $\pi$
according to SU(3). If the phase $\psi_{\rho}$ is
a free  parameter of the fit, its obtained value is consistent
with $\pi$ in agreement with simple quark
model~\cite{NuclPhys}. We fixed $\psi_{\rho}$ at $\pi$ while
determining the $\phi$ meson parameters.
$A_{K^+K^-}$ is a constant complex amplitude describing possible
contributions from excited vector states.
The energy dependence of the total width for a meson $V$ is chosen
as in Ref.~\cite{width}.
The function Z(s) given by the relation
$$
Z(s)  =  1+\alpha\cdot\pi\cdot\frac{1+v^2}{2\cdot v},
$$
$$
v     =  \sqrt{1-\frac{m^2_{K^{\pm}}}{s}}
$$
describes the Coulombian interaction of charged kaons in
the final state~\cite{Colomb}.

Masses and total widths of the $\rho(770)$ and $\omega(782)$
resonances were taken from Ref.~\cite{PDG}.

The product $B(\phi\to e^+e^-)B(\phi\to K^+K^-)$ is related to the
peak cross section $\sigma(\phi\to K^+K^-)$ according
to the formula:
$$
 \sigma(\phi\to K^+K^-) = 12\pi\frac{B(\phi\to e^+e^-)B(\phi\to K^+K^-)}{m^2_{\phi}}
$$
and this parameter along with the $\phi$ meson mass and total width
is determined from the fit:
%The following $\phi$ meson parameters have been obtained from the fit:
$$
\sigma(\phi\to K^+K^-) = 2016\pm8\pm 44\mbox{ nb},
$$
$$
m_{\phi} = 1019.441 \pm 0.008\pm 0.080\mbox{ MeV/c}^2,
$$
$$
\Gamma_{\phi} = 4.24\pm0.02\pm 0.03 \mbox{ MeV}.
$$
And from the other fit, where instead of the peak cross section we have a product
of the branching fractions as a free parameter, we obtain:
$$
B_{\phi\to e^+e^-}\cdot B_{\phi\to K^+K^-} = (14.27\pm0.05\pm0.31)\cdot 10^{-5} ,
$$
where the first error is statistical and the second is systematic.
If we keep $A_{K^+K^-}$ as a free parameter, it is consistent with
zero
and we fixed it at this value while
determining the $\phi$ meson parameters.

The systematic error in $\phi$ meson mass and total width is dominated by the accuracy of
the beam energy determination described in
Ref.~\cite{lum}.

The obtained value of $\sigma(\phi\to K^+K^-)$ agrees
with the results
of CMD-2 \mbox{$2001\pm65\pm82$}~nb~\cite{CMD-295} and
SND \mbox{$1967\pm23\pm140$}~nb~\cite{SND01}
and is more precise.

The values of $m_{\phi}$ and $\Gamma_{\phi}$ obtained in this work
are strongly correlated with the corresponding values
obtained in our analysis of the neutral kaon pair
production in~Ref.~\cite{klks} because they are based on almost the same data
sample and therefore should be not averaged together.
The values of both $\phi$ meson mass and width obtained in this
analysis agree with the world average values and that for the width
is more precise.

%The value of $\sigma(\phi\to K^+K^-)$ should be compared with the results of most
%precise experiments $\sigma(\phi\to K^+K^-) = 1993\pm65\pm82$ (nb)~\cite{CMD-295}
%and $\sigma(\phi\to K^+K^-) = 1967\pm 23\pm 140$ (nb)~\cite{SND01}.

The parameter $B_{e^+e^-}\cdot B_{K^+K^-}$ is in good agreement with
the world average
value~\cite{PDG} $(14.60\pm 0.33)\times 10^{-5}$ and has the same accuracy.

\begin{figure}[tbh]
\includegraphics[width=0.8\textwidth]{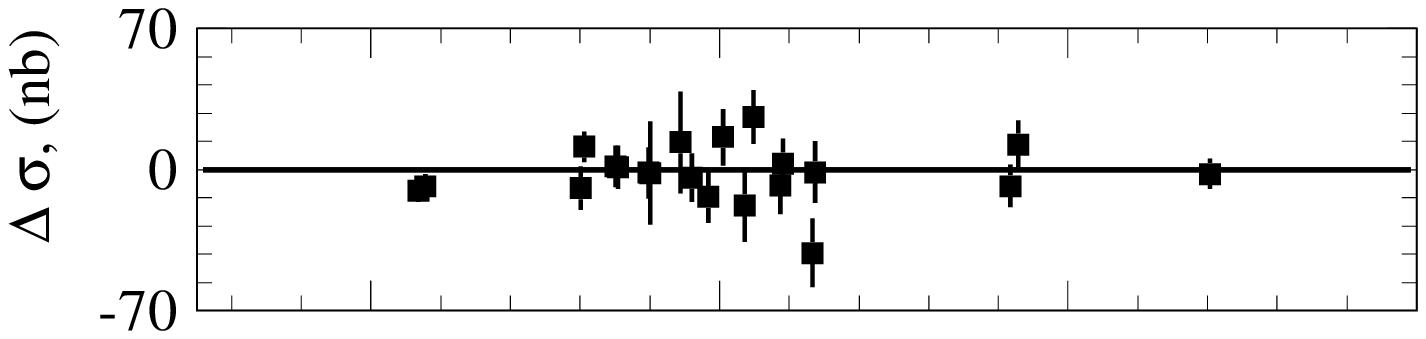}
\includegraphics[width=0.8\textwidth]{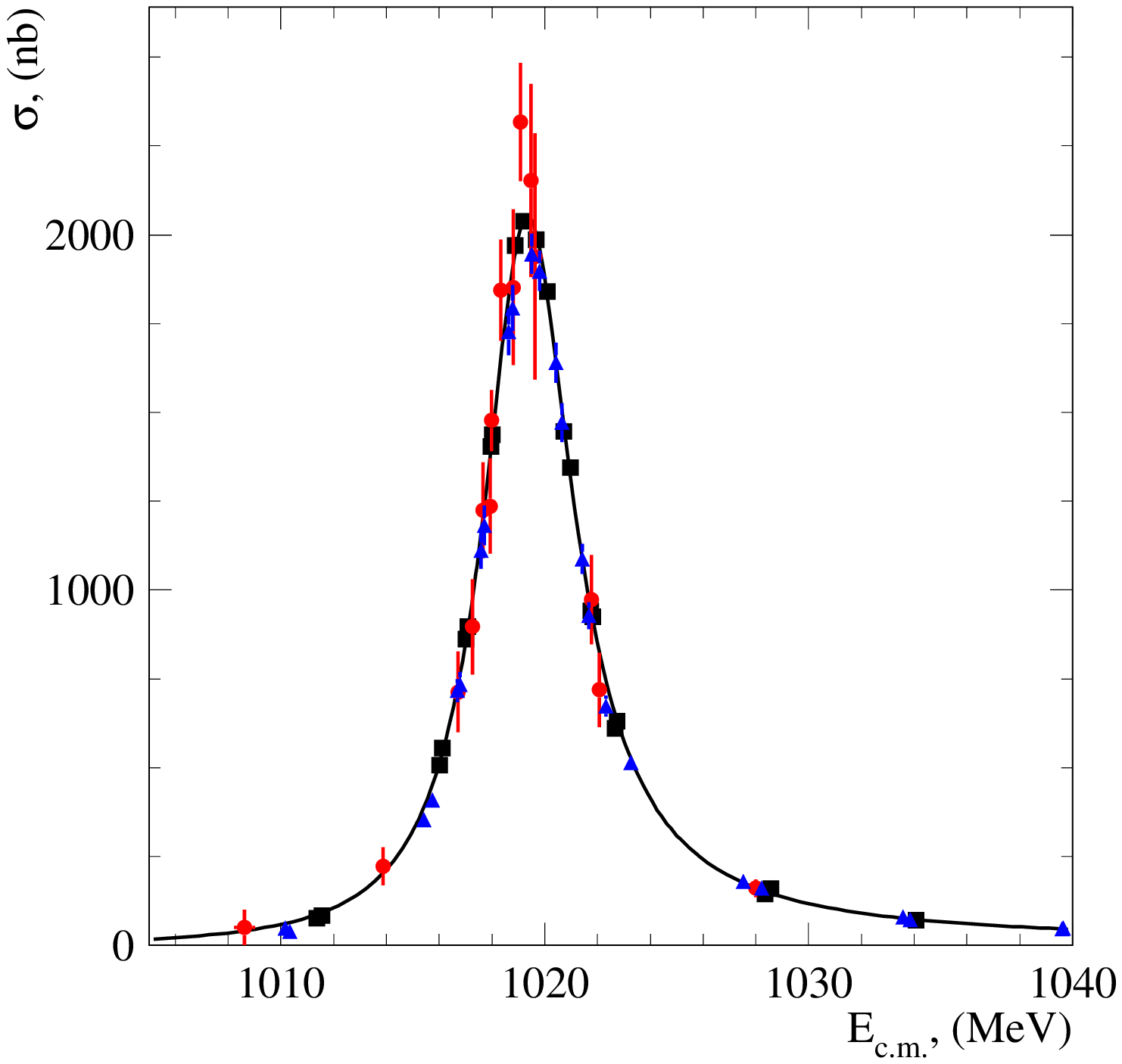}
\caption{\label{phiexps} (Top) The deviations of the measured cross
section from the fitting curve. (Bottom) The experimental cross section of
the reaction $e^+e^-\to\phi\to K^+K^-$ obtained in the present
analysis (squares), earlier CMD-2 experiment~\cite{CMD-295} (circles)
and SND experiment~\cite{SND01} (triangles)}
\end{figure}

In Fig.~\ref{phiexps} we show the energy dependence of the
cross section obtained in this work as well as the results of the
most precise previous experiments~\cite{CMD-295,SND01}.
The results of all experiments are in good agreement.

%of the other
%most precise experiments
%is shown in Fig.~\ref{sig0} and in Table~\ref{tab:sig0}.

%\begin{figure}
%\includegraphics[width=0.8\textwidth]{sigma0.eps}
%\caption{\label{sig0} The value of $\sigma(\phi\to K^+K^-)$, obtained in present work in compare with the
%previous CMD-2 result from~\cite{CMD-295} and with the SND result from~\cite{SND01}}
%\end{figure}

%\begin{table}
%\caption{\label{tab:sig0}Value of $\sigma(\phi\to K^+K^-)$, obtained in present analysis as well as
%in experiments~\cite{CMD-295,SND01}}
%\hspace*{5mm}
%\begin{center}
%\begin{tabular}{|c|c|}
%\hline
%Experiment & $\sigma(\phi\to K^+K^-)$,(nb) \\
%\hline
%\hline
%CMD-2 (1995)~\cite{CMD-295}   & $1993\pm 65\pm 82$  \\
%SND   (2001)~\cite{SND01}     & $1967\pm 23\pm 140$ \\
%CMD-2 (2008), this work       & $2033\pm  8\pm  45$ \\
%\hline
%\end{tabular}
%\end{center}
%\end{table}
%
\section{Discussion}
\hspace*{\parindent}Significant improvement of the
systematic accuracy of the cross section (from 4\% to 2.2\%) is
achieved due to additional analysis of events with only one charged
kaon. It allows to take into account a possible difference
of nuclear interactions, decays in flight and reconstruction
efficiency of the charged kaons in data and MC
simulation.
The trigger efficiency is also extracted directly from the data
and is in good agreement with the MC simulation.

Using this precise measurement we recalculate the contribution of the
reaction  $e^+e^-\to\phi\to K^+K^-$ to the hadronic part of the
theoretical prediction for the anomalous magnetic moment of muon.
It can be calculated via the dispersion
integral~\cite{DispInt}:
\begin{eqnarray}
a^{\rm{had,LO}}_{\mu} & = & \frac{\alpha^2(0)}{3\pi^2}\int\limits_{4m^2_{\pi}}^{\infty}ds\frac{K(s)\cdot R(s)}{s},
\end{eqnarray}
where K(s) is the QED kernel~\cite{QEDKer}, R(s) denotes the ratio of
the ``bare'' cross-section for $e^+e^-$ annihilation into
hadrons to the  muon pair cross section.
%As it was calculated in~\cite{Davier}
%$a^{had,LO}_{\mu}$ at $\phi$ ($\sqrt{s}$ = 1.0~--~1.055 GeV) is
%equal to $a^{had,LO}_{\mu} = (35.71\pm 0.84\pm 0.20)\cdot 10^{-10}$
%and about 50\% from this value is the contribution from $K^+K^-$ channel.
Using data on the $e^+e^-\to\phi\to K^+K^-$ cross section
from~\cite{CMD-295,SND01} one obtains the following average $K^+K^-$
contribution to $a^{\rm{had,LO}}_{\mu}$
in the c.m. energy range $\sqrt{s}$ =  1.011~--~1.055 GeV:
(15.28$\pm$0.16$\pm$0.78)$\cdot$10$^{-10}$. From the results of
the present work the new value of the $K^+K^-$ contribution
to $a^{\rm{had, LO}}_{\mu}$ in the same energy range is
(15.53$\pm$0.15$\pm$0.33)$\cdot$10$^{-10}$. It agrees with the
previous one and is more precise.

\section{Conclusions}
\hspace*{\parindent}Using a data sample of 5.42$\times$10$^5$
reconstructed events with one or two reconstructed charged kaons
collected at CMD-2, the most precise measurement of the
cross section of the reaction $e^+e^-\to\phi\to K^+K^-$
has been performed.
The estimated systematic error in the cross section is 2.2\%.
The following $\phi$ meson parameters have been determined:
\begin{center}
$\sigma(\phi\to K^+K^-) = 2016\pm 8\pm 44$~nb, \\
$m_{\phi} = 1019.441 \pm 0.008\pm 0.080$~MeV/c$^2$, \\
$\Gamma_{\phi} = 4.24\pm0.02\pm 0.03$~MeV,\\
$B_{ee}\cdot B_{KK} = (14.27\pm0.05\pm0.31)\cdot 10^{-5}$.
\end{center}
The obtained results agree with and are more precise than the results
of other measurements.

\section{Acknowledgements}
\hspace*{\parindent} The authors are grateful to the staff of VEPP-2M for
excellent collider performance, to all engineers and technicians who
participated in the design, commissioning and operation of CMD-2.
\par This work is supported in part by the grants \mbox{INTAS YSF 06-100014-9464},
\mbox{INTAS 1000008-8328}, \mbox{RFBR 06-02-16156},\mbox{RFBR 06-02-26590}, \mbox{RFBR 06-02-16445}.

\end{document}